\documentclass[]{raa}
\usepackage{graphicx,times}
\usepackage{natbib}
\usepackage{amssymb,amsmath}
\bibpunct{(}{)}{;}{a}{}{,}

\usepackage{hyperref}
\hypersetup{pdftitle = The title of my PDF, pdfauthor = My name, pdfsubject= The subject, pdfkeywords = keyword1 keyword2 keyword3}
\hypersetup{colorlinks = true, linkcolor = green, anchorcolor = red, citecolor = blue, filecolor = red, pagecolor = red, urlcolor = red}

\newcommand{\nlogn}{~${\mathcal O}(N\log{N})$~} 

\newcommand{\Code}{photo${N}$s-2~}
\newcommand{\erfc}{\textrm{erfc}}

\begin{document}
\title{A Hybrid Fast Multipole Method for Cosmological $N$-body Simulations}

\setcounter{page}{1}

\author{Qiao Wang}

\institute{
Key Laboratory for Computational Astrophysics, National Astronomical Observatories, Chinese Academy of Sciences, Beijing 100101, China\\
School of Astronomy and Space Science, University of Chinese Academy of Sciences, Beijing 100049, China\\
\vs \no
{\small Received 0000 xxxx 00; accepted 0000 xxxx 00}
}

\abstract{
We investigate a hybrid numerical algorithm aimed at the large-scale cosmological $N$-body simulation for the on-going and the future high precious sky surveys. It makes use of  a truncated Fast Multiple Method (FMM) for short-range gravity, incorporating with a Particle Mesh (PM) method for long-range potential, which is applied to deal with extremely large  particle number. In this work, we present a specific strategy to modify a conventional FMM by a Gaussian shaped factor and provide quantitative expressions for the interaction kernels between multipole expansions. Moreover, a proper multipole acceptance criteria for the hybrid method is introduced to solve potential precision loss induced by the truncation. Such procedures reduce the mount of computation than an original FMM and decouple the global communication. A simplified version of code is introduced to verify the hybrid algorithm, accuracy and parallel implementation.
}

\authorrunning{Q. Wang}  
\titlerunning{}
\maketitle

\section{Introduction}
In the early Universe, extremely hot and dense baryons and photons are strongly interacting prior to the recombination, The relic of the fluctuation of photons is imprinted on the Cosmic Microwave Background Radiation which can be detected at the radio band, as the Universe is expanding\citep{2018arXiv180706209P}. Meanwhile the initial fluctuation of the mass is growing up due to the gravitational collapse to form planets, galaxies, cluster haloes and large scale structure of the universe at present\citep{1970ApJ...162..815P}. Several primary cosmic probes can extract the information locked in the mass distribution, such as mass function of cluster counting, the sound horizon of Baryonic Acoustic Oscillation as a standard ruler, power spectrum of cosmic gravitational lensing, etc\citep{1987MNRAS.227....1K,2005ApJ...633..560E}. The evidence combined with various data from sky surveys\citep{2010ApJ...708..645R, 2018MNRAS.480.3879A} supports a modern picture of cosmology with two mysterious components, dark matter and dark energy, whose natures are still puzzles of standard physics\citep{2019PhRvL.122q1301A}. 

The next generation sky surveys will reveal the dark side of the Universe by distant galaxies and quasars as cosmic probe tracers, including DESI\footnote{https://www.desi.lbl.gov/}, EUCILD\footnote{https://www.euclid-ec.org/}, LSST\footnote{https://www.lsst.org/}. One crucial step is to generate a simulated catalogue of those tracers. However, it is not trivial to produce the related observable for galaxy formation involves complicated astrophysical processes and nonlinear evolution\citep{2000MNRAS.319..168C, 2002ApJ...575..587B, 2016MNRAS.456.4156K, 2011MNRAS.413..101G}. The mock tracers, such as the emission line galaxies resided at the smaller structures and earlier epoch, require underlying cosmological simulations with unprecedented resolution and volume to understand their mask effects, cosmic variance, redshift uncertainties, and etc for the next generation surveys. Various techniques are developed for large-scale simulations carried out on top supercomputers with trillion particles.
 
The movement of particles in a expanding background can be modelled by equivalent Newtonian gravity with a periodic boundary conditions. One natural solver of this model is referred as to Particle-mesh (PM) method\citep{1988csup.book.....H} by convolution of Green function of gravity. The convolution has cost of \nlogn, calling Fast Fourier Transformation (FFT) by two times.  However, PM can only deal with the scale above the computing grid so that it needs a compensated sub-grid gravity solver, such as a truncated Particle-Particle (PP) direct summation. The pioneer cosmological simulation based on PP+PM method, so called P$^3$M, identified the structure of the cosmic web\citep{1985ApJS...57..241E, 2002ApJ...574..538J}. Once the system has apparently condensed, it will fail to reduce to ${\mathcal O}(N^2)$ due to the domination of PP interactions.

Another solver with cost of \nlogn is introduced by \citet{1986Natur.324..446B}. They make use of octal tree to organize the particles. The detail of source cells could be negligible, since it attracts a particle well-separated like the gravity of a mass points. Thus, any particle just concerns on the cells or particles ``near enough", determined by a opening angle to control the precision of the acceleration. \citet{2013arXiv1310.4502W} modified a tree code (2HOT) to run on Graphics Processing Unit (GPU) and \citet{Bedorf:2014:PGT:2683593.2683600} optimized ${\it Bonsai}$ \citep{2013CoPhC.184..456P} to achieve performance of 24.77 Pflops for the Milky Way simulation.

Tree code is less sensitive to the particle clustering than a PM code but PM method is stable and rapid for a regular and periodic mesh. A hybrid TreePM method can combine the merits from both of methods\citep{2002JApA...23..185B,2003NewA....8..665B}. The Millennium simulations\citep{2005Natur.435..629S} is carried out with over $10^{10}$ particles by the parallel TreePM code of Gadget-2\citep{2005MNRAS.364.1105S} over ten years ago. The idea of TreePM is also effective on heterogeneous platform, such as HACC \citep{2016NewA...42...49H, 2013hpcn.confE...6H}. Recently, \citet{Ishiyama:2012:PAN:2388996.2389003} run a trillion particles cosmological simulation on K computer.

To face the challenge of high precision, the scale of simulation is increasing as the capability of the supercomputers. That requires faster algorithm to deal with a unprecedented particle number. Fast Multipole Method (FMM) has a nearly linear computational complexity of ${\mathcal O}(N)$. It is originally introduced by \citet{1987JCoPh..73..325G}. \citet{1999JCoPh..155..468C} extends it into three dimension. Similar with Tree method, FMM also builds a tree but computes the interactions between cells. The early FMM works on spherical polar coordinate. It is also successful to be accelerated on GPU devices, such as ExaFMM\footnote{http://www.bu.edu/exafmm/} \citep{Gumerov:2008:FMM:1399648.1399820, 2011arXiv1110.2921Y}. Its precision is controlled by the order of expansions. Moreover, an implementation in a Cartesian coordinate may be more suitable to astrophysical simulation with a acceptance criteria based on opening angle, instead of ``children of parent's brother"\citep{2000ApJ...536L..39D, 2002JCoPh.179...27D}. For some mass dependent criteria, its complexity can approach ${\mathcal O}(N^{0.86})$ by using a dual tree traversal. Recently, \citet{2017ComAC...4....2P} reports that the application of cosmological simulation is able to involve 2 trillion particles by software PKDGRAV-3 on supercomputer $Piz~Daint$.

This paper is organized as follows. In section \ref{sec:method}, we briefly introduce the fundamental approach on FMM then we investigate the combination of FMM and PM method, two current  methods, for the cosmological simulation. The detail of our algorithm will be discussed in section \ref{sec:pmfmm}. A corresponding Multipole Acceptance Criteria (MAC) is taken into account in section \ref{sec:mac}. We measure the reduction of kernel interactions via the hybrid method than a conventional one in section \ref{sec:imp}. In section \ref{sec:ph2}, we introduce a parallel implementation to verify the algorithm. Finally, we summarize in the last section.

\section{Algorithm}
\label{sec:method}
\subsection{Fast Multipole Method}
\begin{figure}[htbp]
\centering
\includegraphics[width=0.7\linewidth]{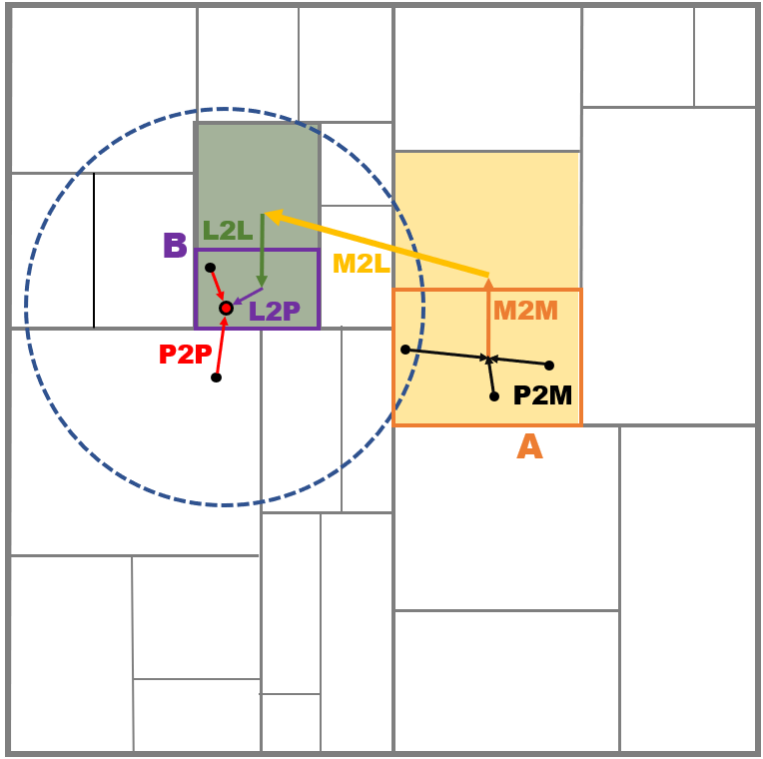} 
\caption{Schematic diagram for FMM. The gravitational potential at certain point (red point with black circle) in Region A is generated by a group of particles in region B. The multipole expansion is computed by P2M in Region A. Then the information is successively propagated via interaction kernels, M2M, M2L, L2L and L2P. The gravity from the nearby particles can be directly accumulated by P2P. The blue dashed circle denotes the cutoff radius.}
\label{fig:pmf}
\end{figure}

In FMM, all particles are organized into a tree and the finest tree cells (or tree nodes) are always a series of particle packs, which are also referred as to leaves. In Fig.~\ref{fig:pmf}, we employ an ORB (Orthogonal Recursive Bisection) tree and set a maximum limiter for particle number in leaves. The particles including parent cell are almost equally divided into two offspring cells down to leaves. 

The gravitational potential at point in region $B$ induced by the particles of region $A$ can be estimated via a series of interaction kernels between the multipole coefficients. We follow the mathematical notations in \citet{2014ComAC...1....1D} and summarize the relevant equations of FMM method in Cartesian coordinates. The multipole coefficients ${\mathcal M}$ of region $A$ is computed by the particles
\begin{equation}
{\mathcal M}_{\mathbf m}({\mathbf z}_A) = \sum_{a \in A} \mu_a \frac{(-1)^m}{{\mathbf m}!}( {\mathbf x}_a - {\mathbf z}_A )^{\mathbf m}, 
\label{eq:p2m}
\end{equation}
where ${\mathbf z}_{A}$ is the geometric (or mass) center of region $A$, $\mu_a$ is mass of the particle labeled by $a$ at the position of $x_{a}$ and the integer vector ${\mathbf n}$ denotes $(n_x, n_y, n_z)$. One can shift a multipole expansion form ${\mathbf z}$ to ${\mathbf z} + {\mathbf x}$ by the summation of
\begin{equation}
{\mathcal M}_{\mathbf m}({\mathbf z}+{\mathbf x}) = \sum^{{\mathbf m}}_{{\mathbf n}=0}\frac{{\mathbf x}^{\mathbf n}}{{\mathbf n}!}{\mathcal M}_{{\mathbf m}-{\mathbf n}}({\mathbf z}).
\label{eq:m2m}
\end{equation}
Given a Green function of $\psi( {\mathbf z}_b - {\mathbf z}_a )$, the coefficients of local expansion ${\mathcal L}$ of potential ${\Psi}$ at the center of region $B$ of ${\mathbf z}_{B}$ is determined by the ${\mathcal M}$ at ${\mathbf z}_a$ by equation 
\begin{equation}
{\mathcal L}_{\mathbf n}({\mathbf z}_B) = \sum_{|{\mathbf m}|=0}^{p-|{\mathbf n}|}{\mathcal M}_{\mathbf m}({\mathbf z}_A){\mathcal D}_{{\mathbf n}+{\mathbf m}}({\mathbf z}_B-{\mathbf z}_A),
\label{eq:m2l}
\end{equation}
where ${\mathcal D}_{\mathbf n} \equiv \nabla^{\mathbf n} {\psi}$ is a ${\mathit traceless}$ operator. For Newtonian gravity or static electricity force, it can be expressed by a traceless displacement tensor $\bar{{\mathbf r}}^{\mathbf n}$ multiplied by the prefactors of 
\begin{equation}
\tilde{f}_{(n)}(r)= (-1)^{n}\frac{(2n-1)!!}{r^{2n+1}}. 
\label{eq:pref}
\end{equation}
Similarly, the multipole ${\mathcal L}$ can be shifted by
\begin{equation}
{\mathcal L}_{\mathbf n}({\mathbf z}+{\mathbf x}) = \sum_{|{\mathbf m}|=0}^{p-|{\mathbf n}|}\frac{{\mathbf x}^{\mathbf m}}{{\mathbf m}!}{\mathcal L}_{{\mathbf m}+{\mathbf n}}({\mathbf z}),  
\label{eq:l2l}
\end{equation}
Thus, the potential at ${\mathbf x}_b$ is approximated by
\begin{equation}
{\Psi}({\mathbf x}_b) = \sum_{|{\mathbf n}|=0}^{p} \frac{1}{{\mathbf n}!} {\mathcal L}_{\mathbf n}({\mathbf z}_B) ( {\mathbf x}_b - {\mathbf z}_B )^{\mathbf n}.
\label{eq:l2p}
\end{equation}

For short, The multipole expansion for the source is labeled by 'multipole' and the expansion of at sink (or target) place is labeled by 'local'. Then the abbreviations of the kernels are as follows: particle-to-particle is referred as to P2P, particle-to-multipole is P2M, multipole-to-multipole is M2M,  multipole-to-local is M2L, local-to-local is L2L, and local-to-particle is L2P, respectively. the multipole and local expansion coefficients are included in all tree cells.

The above equations can be utilized to transmit the information of gravity from a particle group to another. The gravity of a certain particle in the purple sink (targeted) leaf induced by the orange source leaf can be computed through a series of successive kernels. The multipole coefficients ${\mathcal M}$ of a (orange) leaf are determined by the discrete mass points in the cell (by using Eq.~\ref{eq:p2m}), which describe mass distribution in source cells or leaves. Then the multipole expansion coefficients in a parent cell are computed by its offspring cells recursively (using Eq.~\ref{eq:m2m}). This procedure is also referred as to upward pass.

The kernel M2L computes the local expansion coefficients of gravitation potential from the multipole of source cells by Eq.~\ref{eq:m2l}. When two cells are well separated each other, the M2L kernel is called to compute the the interactions from yellow to green cells in this illustration. The local expansion coefficients are passed level-by-level downward (using Eq.~\ref{eq:l2l}) till a leaf is met. The gravity and potential of the targeted particles (red point with black circle) is calculated by the local expansion about the center of purple leaf, using Eq.~\ref{eq:l2p}. 

The nearby particles surrounding the target are too close to be dealt with by above kernels and they must be directly accumulated their interaction. Since a direct N-body summation is ${\mathcal O}(N^2)$, it can be quite time consuming. The procedure is referred as to P2P, which actually can be considered as interactions between particle packs in our implementation as well.

\subsection{Combination of Particle-Mesh and Fast Multipole Method}
\label{sec:pmfmm}
Analogue to TreePM, the essence of Particle-Mesh and Fast Multipole Method (PM-FMM) is also to split gravity into two parts by scale. A truncated short-range part of gravity is computed by FMM and a smoothed long-range part is by PM method. A combination of two parts must be equivalent with an original inverse-square law at the split scale, by fine tuning its splitting function. \citet{2002JApA...23..185B} suggests a Gaussian form as transition function for TreePM and we generalize it to the PM-FMM in this section. 

Specifically, the convolution for PM with a grid size of $\Delta_{g}$ needs an Gaussian function $\exp(-{k^2}/{4k_s^2} )/\sqrt{\pi}$ as a filter to suppress the Green function of Poisson equation, where $k_s$ is the wave number of split scale of $r_s \sim 1.2 \Delta_{g}$. 

\begin{figure}[htbp]
\centering
\includegraphics[width=0.7\linewidth]{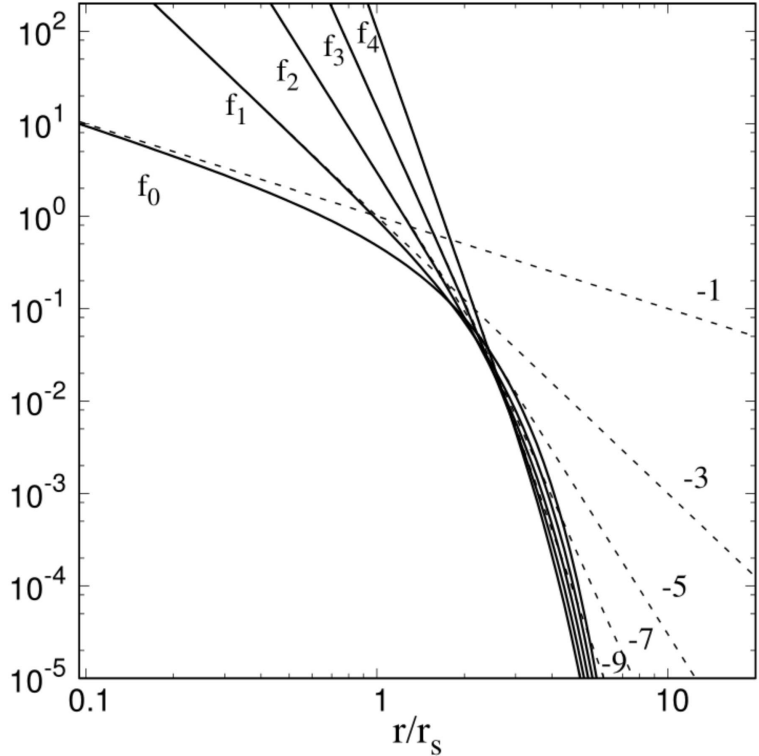}
\caption{Reduction effect. All prefactors are truncated at the cutoff radius $\sim 5 {r_s}$.}
\label{fig:fac}
\end{figure}
Correspondingly, the short range gravity must be properly modified to compensate the underestimation from PM method. As a result, the potential $r^{-1}$ is modified by error function as
\begin{equation}
\phi({r})=f_{(0)}({r}) = \frac{1}{r} \erfc \left( \frac{r}{2r_s} \right).
\end{equation}
and an inverse-square law is modified by
\begin{equation}
{\mathrm g}({r})=f_{(1)}{\mathbf r} = -\frac{\mathbf r}{r^3} \left[ \erfc \left( \frac{r}{2r_s} \right) + \frac{r}{r_s\sqrt{\pi}} \exp \left( -\frac{r^2}{4r_s^2} \right) \right].
\end{equation}
Moreover, the operator ${\mathcal D}_{\mathbf n}=\nabla^{\mathbf n} [\erfc(|{\mathbf r}|/r_s) {\phi}_{N}] $ for a truncated potential is employed for computation of FMM, where $ {\phi}_{N} $ is Newtonian potential.

For the higher orders, the M2L kernel in truncated algorithm can be implemented by minimum substitutions of the original prefactor $\tilde{f}_{(n)}$ (Eq.~\ref{eq:pref}) by the following ones without tildes:
\begin{equation}
\begin{aligned}
f_{(2)}({r}) &= \frac{3}{r^5}  \erfc \left( \frac{r}{2r_s} \right) +\frac{1}{\sqrt{\pi}}\exp \left( -\frac{r^2}{4r_s^2} \right) \\
&\times \left[  \frac{3}{r_s r^4}  + \frac{1}{2r_s^3r^2} \right], 
\notag
\end{aligned}
\end{equation}
\begin{equation}
\begin{aligned}
f_{(3)}({r}) &= -\frac{15}{r^7}  \erfc \left( \frac{r}{2r_s} \right) - \frac{1}{\sqrt{\pi}} \exp \left( -\frac{r^2}{4r_s^2} \right) \\
&\times  \left[   \frac{15}{r_sr^6} + \frac{5}{2r_s^3r^4}+ \frac{1}{4r_s^5r^2}  \right],   \notag
\end{aligned}
\end{equation}
\begin{equation}
\begin{aligned}
f_{(4)}({r}) &= \frac{105}{r^9}  \erfc \left( \frac{r}{2r_s} \right) + \frac{1}{\sqrt{\pi}} \exp \left( -\frac{r^2}{4r_s^2} \right) \\
& \times  \left[ \frac{105}{r_sr^8} +\frac{35}{2r_s^3r^6} + \frac{7}{4r_s^5r^4} + \frac{1}{8r_s^7r^2} \right] .
\notag
\end{aligned}
\end{equation}

 Fig.~\ref{fig:fac} demonstrates the comparison of two kinds of prefectors. It is apparent that all interaction are truncated at the radius of $5 \times r_s$,  a cutoff radius $R_{\rm cutoff}$, so that the gravity of short range is negligible beyond that scale. For P2P kernel, only $f_{(1)}$ is needed.

One can compute the prefactor of any order $p$ by equation of
\begin{equation}
\begin{aligned}
(-1)^{p}r_s^{2p+1}f_{(p)}(x) &= \frac{(2p-1)!!}{x^{{2p+1}}}  \erfc \left( \frac{x}{2} \right) \\
&+ \sum_{q=1}^{p}  \frac{2^{q-p}(2p-1)!!}{(2p-2q+1)!!}\frac{e^{ -{x^2}/{4} } }{\sqrt{\pi}x^{2q}},
\end{aligned}
\end{equation}
where ${x \equiv {r}/{r_s}}$. Or a more effective approach to calculate all of prefactors in successive orders via a recurrence form of
\begin{equation}
-r_s^2x^2 f_{(p+1)}({x})= {(2p-1)} f_{(p)}(x) + \frac{e^{ -{x^2}/{4} } }{2^{p-1}\sqrt{\pi}r_s^{2p-1}}.
\end{equation}

The hybrid algorithm is illustrated in Fig.~\ref{fig:pmf}. The dash circle denotes a cutoff radius. The contribution from FMM is localized within the cutoff radius.

\subsection{Multipole Acceptance Criteria \& Error Estimation}
\label{sec:mac}
A dual tree traversal method can complete all of kernel computation by once tree walking. For local tree, the traversal begins with root-root pair but it can begin with any pair of cells. If two cells are "well separated", a M2L kernel will be employed to compute the local multipole of sink cell induced by the source one. Otherwise one of two cells, usually a larger one, needs be opened. The traversal recursively succeeds in the opened cell until walking at end of a tree. The interaction between two "nearby" leaves must use P2P kernel.

\begin{figure}[htbp]
\centering
\includegraphics[width=0.65\linewidth]{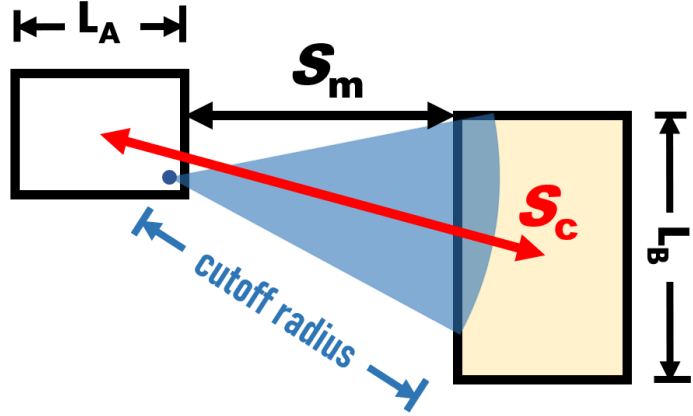}
\caption{Multipole Acceptance Criteria. $L_A$ is side length of sink (targeted) tree cell and $L_B$ is the source one. The red arrow is the separation $S_c$ between the centers of two nodes and  $S_m$  is minimum distance between boundaries of tree cells. Two boxes are still physically relevant despite $S_c$ beyond the cutoff radius.}
\label{fig:mac}
\end{figure}

Therefore the definition of ``well separated" will be influenced by Multipole Acceptance criteria (MAC) . One can define the geometric relation by the parameter of $opening~angle$ $ \theta = L/S$, where $L$ is the length of cell and $S$ is the separation between two cells. It is still various to choice of those lengths. In PKDGRAV3, \citet{2017ComAC...4....2P} makes use of the opening radius $RO=b_{max}/{\theta}$ as a criteria, where the $b_{max}$ is from the mass center of the cell. \citet{2014ComAC...1....1D} introduces to a mass-dependent acceptance criteria, which suppress the cost of ${\mathcal O}(N^{0.86})$  less than a linear complexity. To demonstrate the method in this work, we choose the maximum side length of source cell as $L$, the separation $S_c$ is the distance between the geometric center of two cells and the $S_m$ is minimum distance between two cells (see Fig.~\ref{fig:mac}).

Besides opening angle, our acceptance criteria needs be adjusted for effects caused by truncation and it is summarized into three items:
\begin{itemize}
\item All of interactions between cell-cell, cell-body and body-body are neglected when the minimum separation $S_m$ is larger than the cutoff radius;
\item Call M2L kernel to compute local expansion coefficients while opening angle $\theta \le \theta_{\rm MAC}$. Otherwise, open the larger one of two cells of interest;
\item Enforce to open the cells if the separation is at the range of the transition region, even the 2nd condition is already met, where transition region is defined by  $S_c > R_{\rm cutoff} > S_{m}$.
\end{itemize}
Our opening angle is defined as $\theta \equiv L_B / S_c$, where the $L_B$ is the maximum side length of source cell. It is apparent that neighbor cells must be always opened according to the 2nd item. 

The first item causes an essential improvement of the hybrid method. The additional 3rd item is due to truncated design. Fig.~\ref{fig:fac} shows that the term $r^{-1}$ is dominated beyond the truncated scale but the higher order multipoles still contribute the expansion coefficients of local gravitational field in the traditional FMM. However in our method, the multipoles in any order, here from $f_0$ to $f_4$, are suppressed within the cutoff radius. Therefore no information can be propagated out the cutoff barrier no matter how many orders are considered. The error of short-range gravity also fails to be suppressed by a stricter $\theta_{\rm MAC}$. Because the shadow region in Fig.~\ref{fig:mac} is always ignored and lost due to the modification of prefactors, despite contributes the short-range gravity yet. The 3rd item is motivated to fix it and guarantee the higher multipole can influence the cells beyond the scale of cutoff radius.

\begin{figure}[htbp]
\centering
\includegraphics[width=0.65\linewidth]{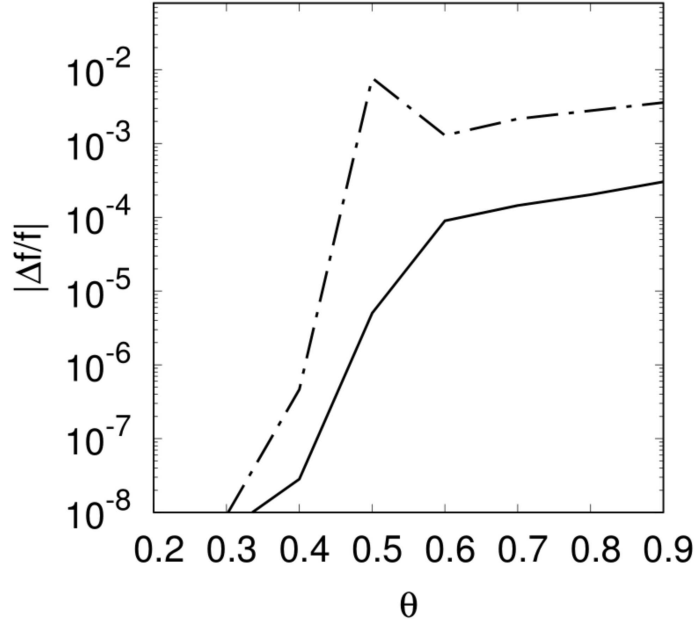}
\caption{Relation of error to opening angle. Relative error of gravity or acceleration is calculated via truncated FMM upto hexadecapole. The solid curve denotes the rms error and dash-dotted curve denotes the maximum error of all particles.}
\label{fig:the}
\end{figure}

Fig.~\ref{fig:the} shows the relative error by the new MAC with the orders up to hexadecapole. The solid curve denotes the root-mean-square (rms) of relative error and the dash-dotted curve denotes the envelop of maximum error. The error of gravity in our criteria with the third item can be well controlled. It drops as the opening angle decreases. But the relation of error to opening angle changes. It breaks the power law of error with $\theta^p$, say $\theta^4$ in hexadecapole, presented in \citet{2014ComAC...1....1D}.

The relation between accuracy and efficiency depends on the choice of control parameters. But it is difficult to derive the optimized parameters from theoretical analysis. The order of FMM can be independently determined, which is constrained by the machine memory. The memory of cells consumes double for every order of FMM adds. Some applications without memory limit employ the FMM upto 10th order with a huge equivalent opening angle. We practically set the order of FMM upto octupole or hexadecapole, since the cosmological simulations are usually pressed for memory. Correspondingly, it keeps sufficient statistics to set the opening angle from 0.3 to 0.4 for regular cosmological simulations. One can decrease the opening angle to improve accuracy and cause a sharp enlargement of the amount of computation. Therefore the parameters must be fine adjusted in accordance with computing power, environments and accuracy tolerance.

\section{Comparison of Computational Cost}
\label{sec:imp}
In the previous section, we describe an algorithm for decoupling long-range and short-range gravity with a proper modification of MAC. Correspondingly, we measure and present that the quantitative reduction of computation amount in this section. There are two extreme density distributions as cosmological evolution. In high-redshift epoch, the density contrast is tiny. the computing box is almost uniformly filled with mass particles. As the evolution of the universe, the initial seed of structure grows up and highly clustering structure are formed, including filaments and the haloes. Most of particles are falling into condensed regions.

\begin{figure}[htbp]
\includegraphics[width=0.45\linewidth]{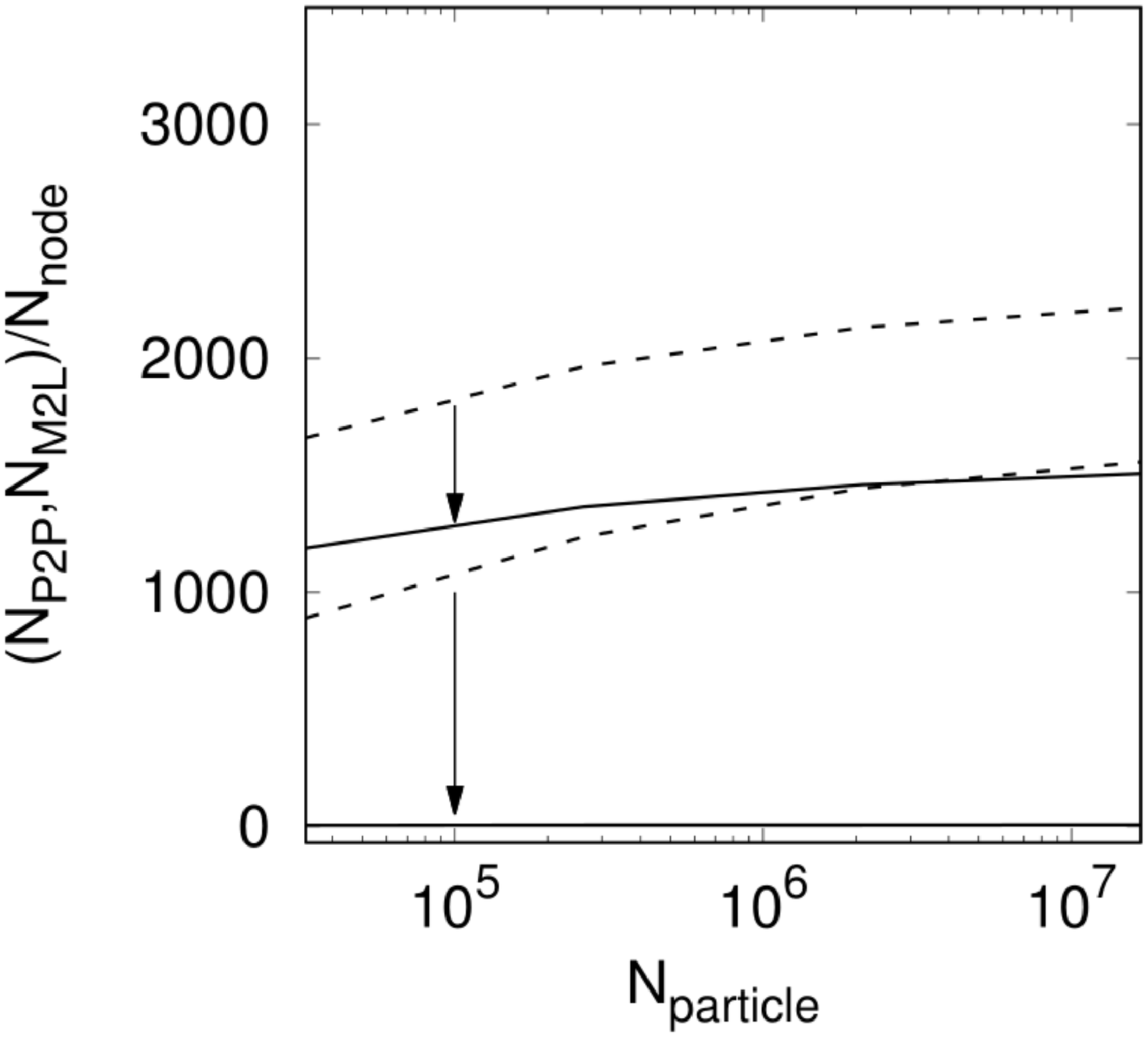}
\includegraphics[width=0.45\linewidth]{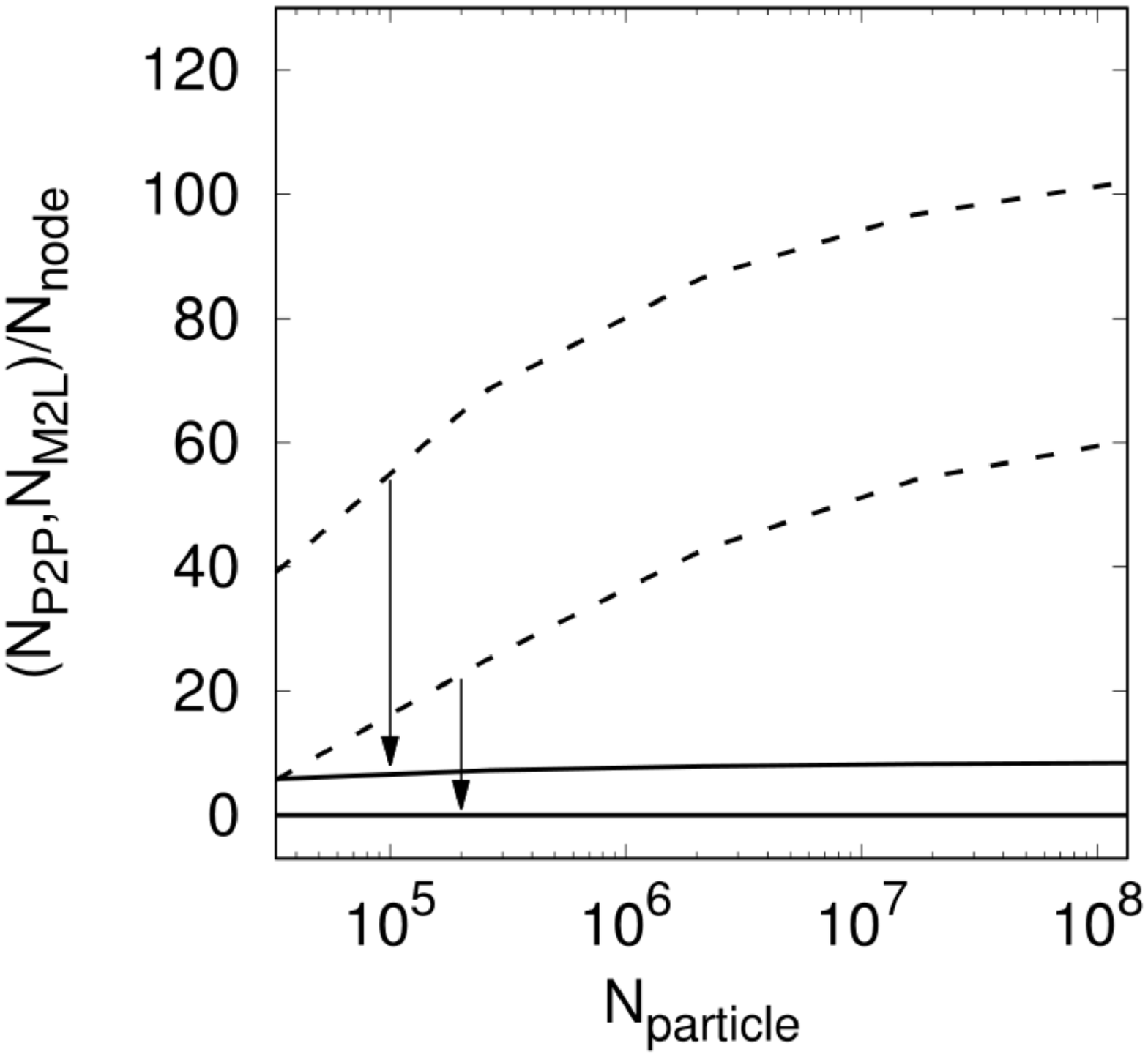}
\caption{Uniformed density distribution. The dotted curves denote the counts of kernel P2P and M2L for a conventional FMM and the solid curves denote truncated FMM in this work. Each leaf (particle pack) only holds one particle in left panel; Each leaf holds 32 particles at most in right panel. Therefore the length of tree cells is shorter than the left panel. }
\label{fig:scun}
\end{figure}

The dashed curves correspond to the conventional case and the solid curves are in this work. Arrows denote the reductions of the amount. the upper curves are for the P2P kernel and lower are M2L. There is apparent improvement, since the mean separation of most cells are beyond the cutoff radius so that the computation of M2L in original FMM vanishes for the uniformed distribution, whose cost is close to P$^3$M method in the uniformed case.

Fig.~\ref{fig:scun} demonstrates the analyses of the uniformed case. Because the most time-consuming kernels of M2L and P2P are directly influenced by the truncated MAC. The run duration depends on the implementation but the count of interaction does not. It only depends on the choice of MAC. In the left panel of Fig.~\ref{fig:scun}, each leaf (particle pack) only holds a particle at most, and 32 particles in the right panel. The count of interaction kernels is determined the depth of tree and spatial configuration. The tree contains less cells in the right panel than right one in the same particle distribution. 

\begin{figure}[htbp]
\includegraphics[width=0.45\linewidth]{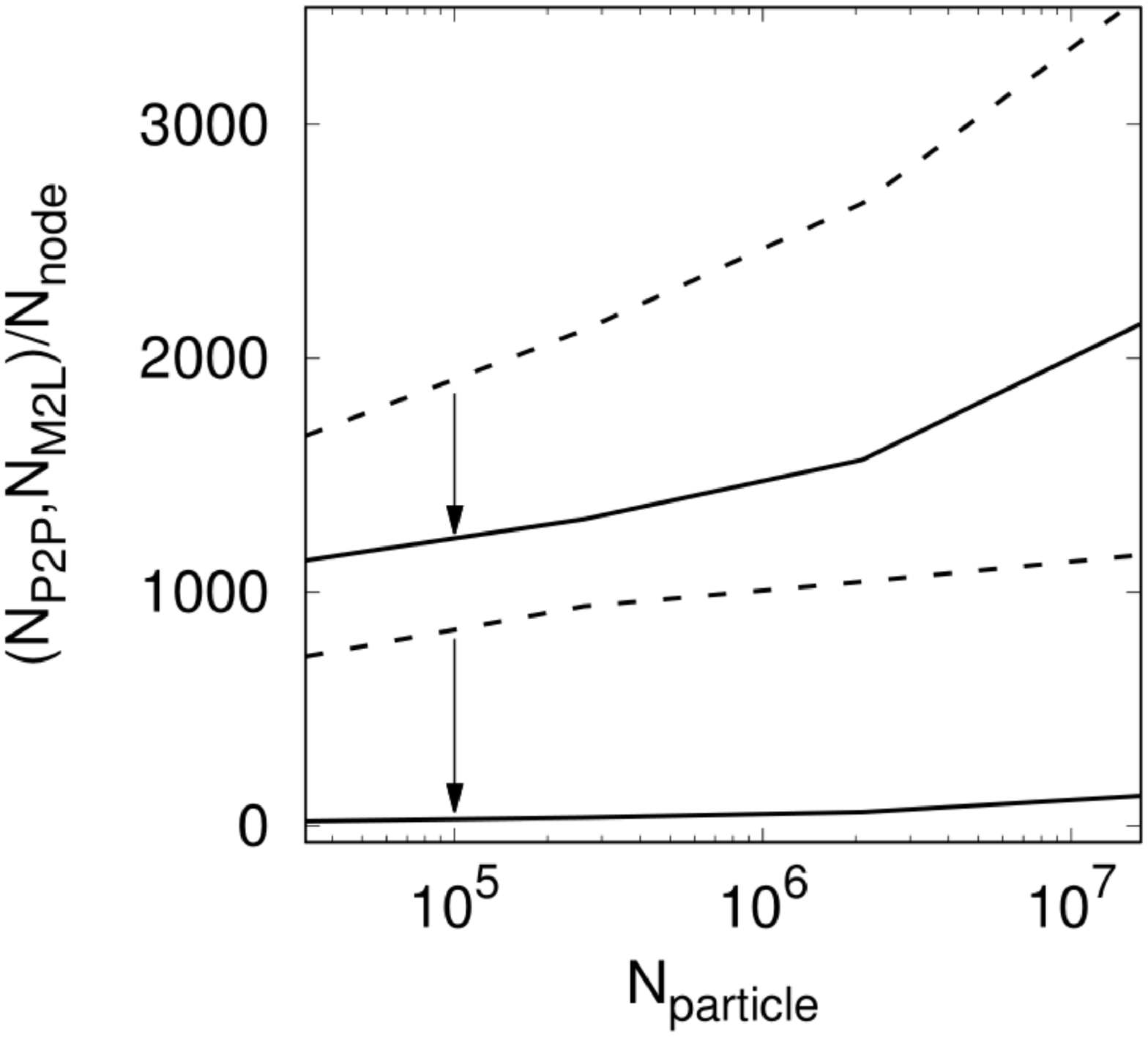}
\includegraphics[width=0.45\linewidth]{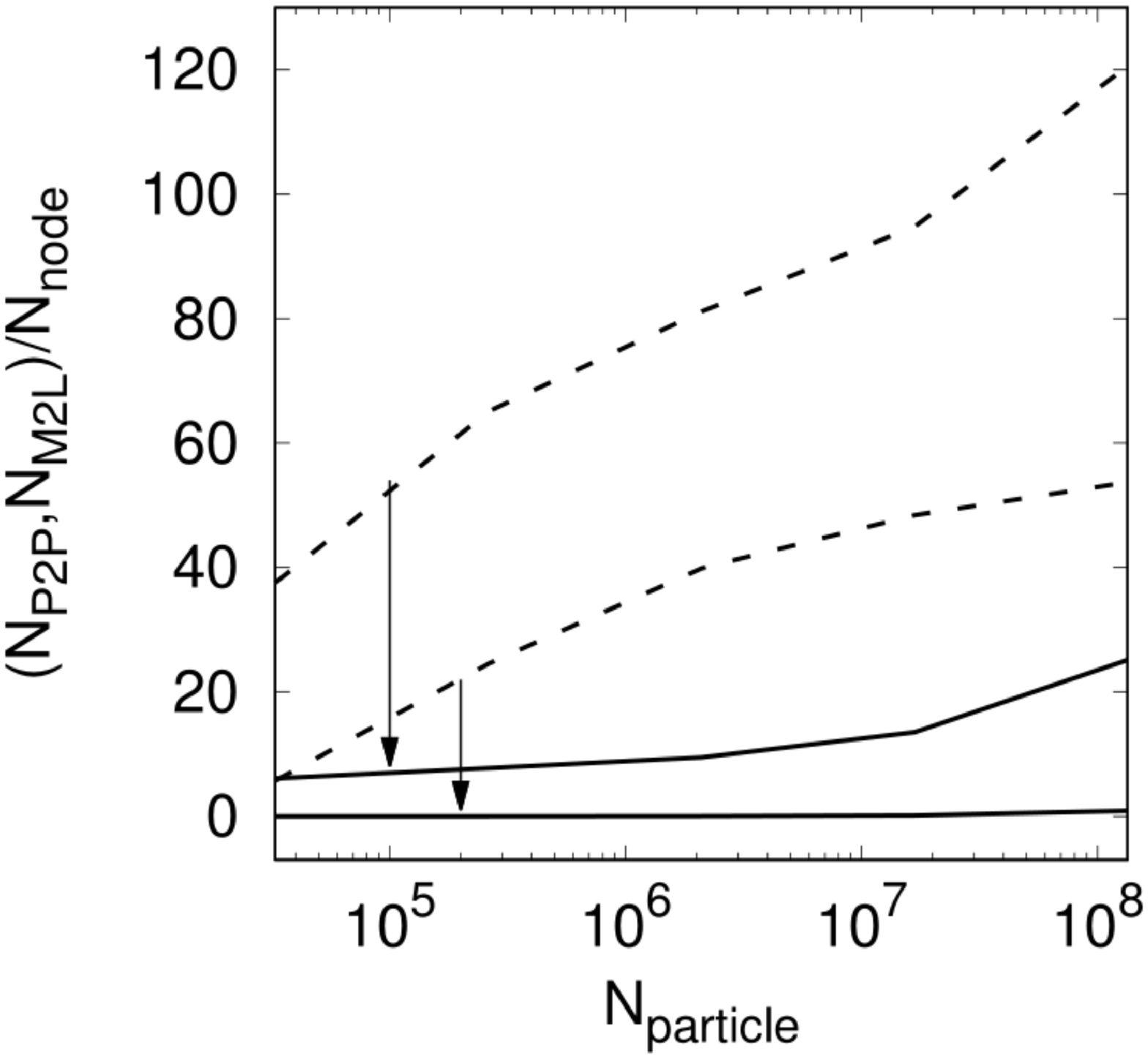}
\caption{Highly clustering density distribution, which is the most condensed case from a $Lambda$CDM cosmological simulation run at redshift $z=0$. The dotted curves denote the interaction counts of kernel P2P and M2L for a conventional FMM and the solid curves denote truncated FMM in the work. Each leaf only contains one particle in left panel; Each leaf contains 32 particles at most in right panel.}
\label{fig:scha}
\end{figure}

Fig.~\ref{fig:scha} demonstrates the similar analyses for the cosmic structure distribution of the universe at present (redshift $z=0$). The notation is the same with Fig.~\ref{fig:scun}. As expected, the total count is larger than the uniformed case, because more particles are constricted into the smaller cells so that more cells are open by the acceptance criteria. The mean separation of cells shrinks below the cutoff radius. but the hybrid method still works to reduce the kernel count. 

Compared with original FMM, kernel interactions of M2L and P2P are reduced in our truncated one. However there exists an additional PM in our method. Theoretically, PM has cost of ${\mathcal O}(N\log N)$, which is worse than FMM. But a PM method can usually be more effectively implemented to save the overall duration. On the contrast, the original FMM makes use of Ewald summation to deal with a periodic boundary condition\citep{2014JCoPh.272..307G}. It is not needed in this hybrid method.

\section{parallel implementation of algorithm}
\label{sec:ph2}
This PM-FMM method is employed by the cosmological N-body simulations code \Code, which is designed for the massively parallel cosmological simulations\footnote{a simplified parallel MPI+openMP version can be download at https://github.com/nullike/photoNs-2.0 to testify the hybrid algorithm.}. Its first version\citep{2018RAA....18...62W} adopts a parallel Tree-PM method and the interactions between particles and tree cells are arranged into a task pool, which is suitable for the optimization, especially on the heterogeneous platform\citep{10.1093/pasj/56.3.521, Hamada:2009:THN:1654059.1654123, 2018RAA....18...62W, 10.1007/978-3-319-93698-7_37}. This second version updates short-range gravitational solver by truncated FMM described in the previous sections ~\ref{sec:method}, instead of the tree method.

\begin{figure}[htbp]
\centering
\includegraphics[width=0.4\linewidth]{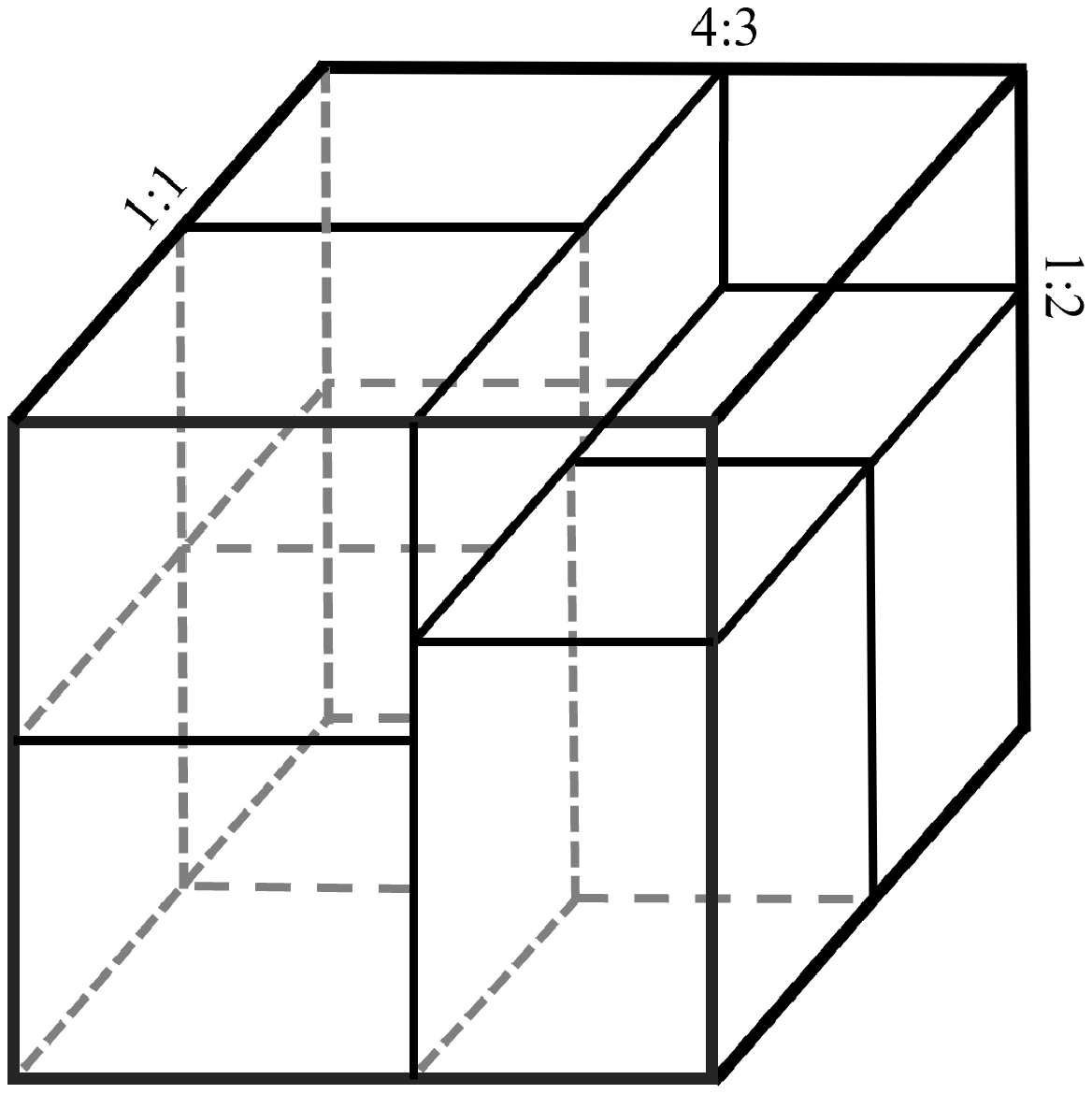}
\includegraphics[width=0.48\linewidth]{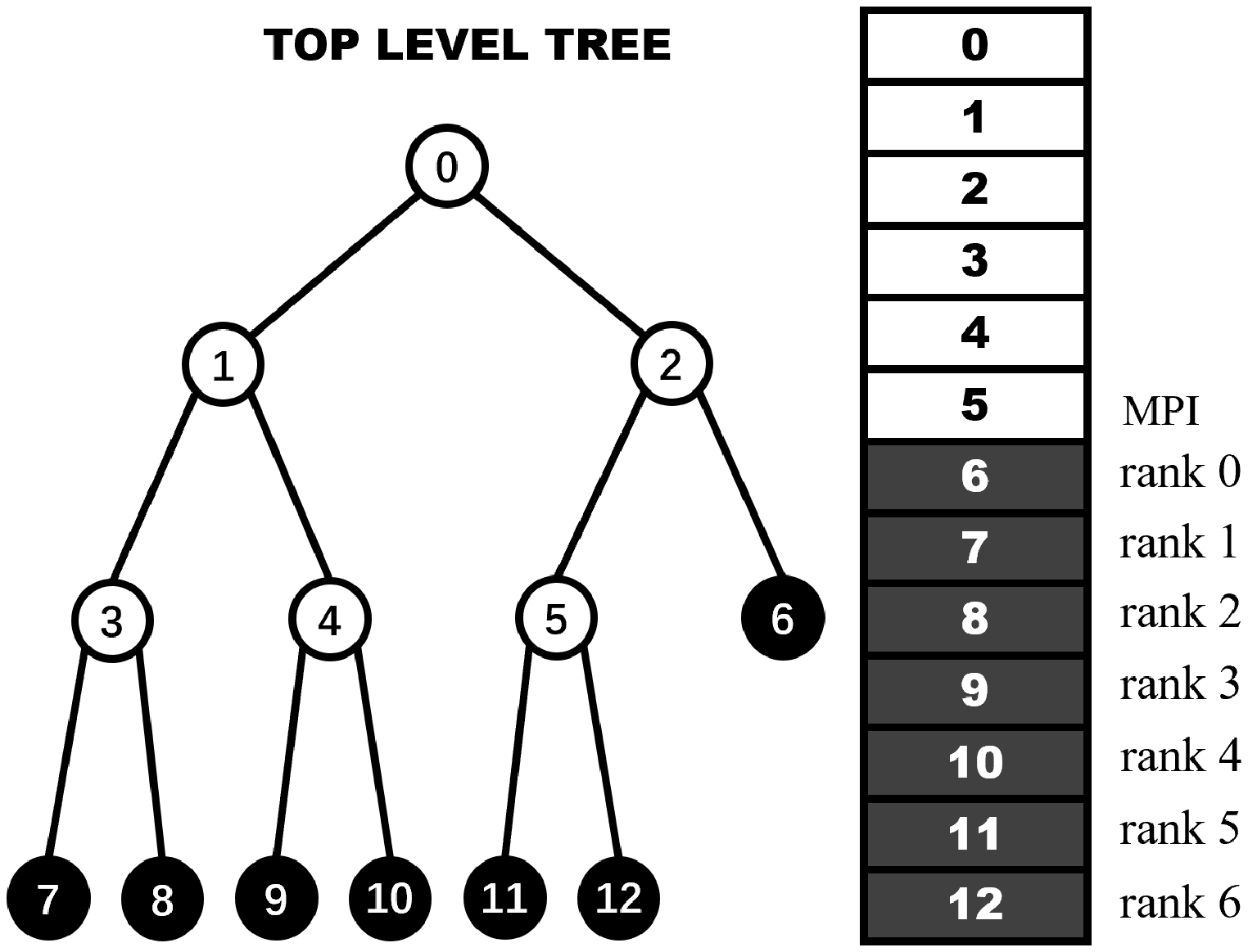}
\caption{The domain decomposition and top level tree.}
\label{fig:toptree}
\end{figure}

In the new version, the domain decomposition returns to an ORB tree across the computing nodes, such as computing sockets or processors with a shared memory. And all of domains and their upper nodes construct a top level tree that needs be stored in all computing nodes. Thus the domain cell is a finest cell in the top level tree but a root cell for local essential tree (LET). The particles are also organized into a $k$-d tree in every computing domain so that a distributed global ORB tree is constructed. Fig.~\ref{fig:toptree} illustrates a domain decomposition for a almost uniformed particle distribution. As an example of 7 processes, the ORB tree firstly distribute the particles along $x$-axe by the fraction of 4:3 in order to balance of particle number, then 4 processes take charge of the left 4/7 volume of box and other 3 processed do the right 3/7, respectively.

The upward pass of P2M and M2M firstly run on the local tree. When all local upward passes are complete, the parallel M2M, M2L and L2L are made for the top level tree. But the inter-domain M2L and P2P still requires the information across the other domains. In \Code, we actively send local tree cells and leaves to the domain may needed it. Using our MAC, a traversal with respect to the closest boundary of target domain will find all of potential cells and leaves involved for the target domain. 
\begin{figure}[htbp]
\centering
\includegraphics[width=0.9\linewidth]{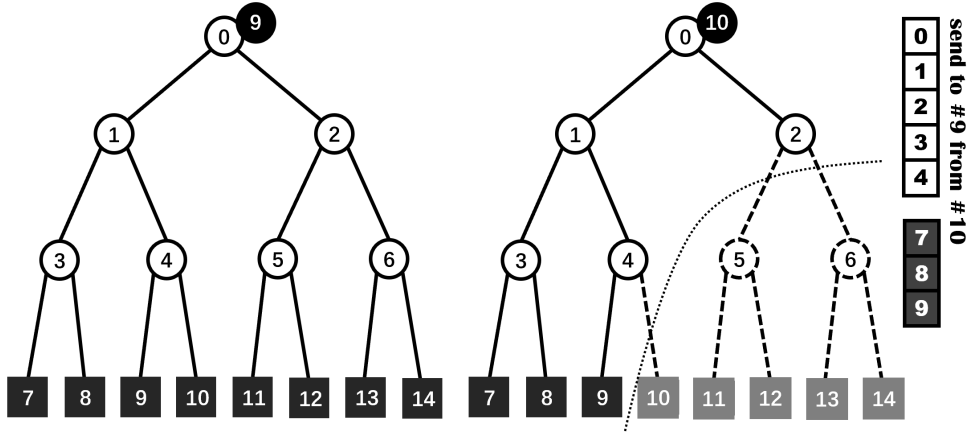}
\caption{The local tree. The tree cells (0-4) and the particle packs (7-9) needs to send to the target domain (Domain 9) in this case.}
\label{fig:loc}
\end{figure}
For instance, there are two local trees which are mounted on domain 9 and 10, respectively in Fig.~\ref{fig:loc}. The cell 2 may be accepted by MAC with respect to "right side" boundary of domain 9 so that nodes 5 and 6 can be safely ignored (see Fig.~\ref{fig:loc}). So are leaves from 11 to 14. The separation between leaf 10 and boundary may be larger than the cutoff radius so that leaf 10 is also ignored. Then a segment of subtree including all necessary information is arranged from the local tree of domain 10. 

Extra memory to $send$ and $receive$ must be allocated to exchange the segment of cells and leaves for the communications. After those parallel operations, local downward pass of L2L and L2P can be executed.

In this version, M2L and P2P operators are most time-consuming two kernels for gravity calculation. For the criteria of ``children of parent's brother", the operation number of M2L and P2P only depends on the length of the tree. But for the MAC in our implementation, it also depends on the clustering of particles. We also make use of a double-buffering task pool to improve the concurrency, for P2P and M2L. As the mass particles collapse into the potential well, the density contrast in the simulation box is wildly different from place to place. Therefore we estimate the work load for single domain by counting the total number of M2L and P2P operations to determine how to redistribute the particles in the next time step. It is similar with the strategy of the code GreeM\citep{2009PASJ...61.1319I, Ishiyama:2012:PAN:2388996.2389003}. Such a feedback strategy usually can control the workload imbalance within $15\%$.

Practically, this hybrid method is designed for the massively parallel on supercomputers with over $10^4$ computing nodes or sockets. Its PM method needs to call FFT subroutine by two times at every single synchronized time step. A conventional FFT library, such as FFTW\citep{Frigo05thedesign}, decompose a mesh into a series of slices along a certain direction. It fails if the number of processes is larger than the side number of mesh. But it exactly happens in a cosmological simulation. In this version, we employ a Fortran library with a pencil decomposition, 2DECOMP\&FFT \citep{20102decomp}, for the PM. As a test, a simulation with $\sim 5000^3$ grid is carried out by over 20,000 processes.

\section{Summary and Discussion}
In this paper, we investigate a hybrid method for the massive application of Cosmological simulations. At the epoch of precious cosmology, FMM with complexity ${\mathcal O}(N)$ is key method to run high resolution simulations on the supercomputers and a traditional PM method still contributes on decoupling the gravity and dealing with the periodic boundary condition. The hybrid algorithm of FMM with PM keeps the benefit of gravity splitting and decrease the amount of computations.

Specifically, we modified the operators of the truncated FMM for the short-range gravity and provide a general form to compute the prefactor of multipoles. We focus on a Gaussian-type truncation. Because its sharp splitting is proven by the TreePM method. In principle, one can choose another truncated function, instead of the exponential form. A polynomial function can be calculated more efficiently for the numerical mathematical library than an exponential one. The modifications of their prefactors can be generated by a similar procedure. The method in this work is different from the Particle mesh multipole method (PMMM), which calls $(p+1)^2$ FFT to directly compute the multipole coefficients\citep{2014arXiv1409.5981N} or Fourier transform on multipoles (FFTM) method\citep{2004IJNME..61..633O}. We do not use Fourier Transformation to calculate the coefficients of multipole expansion but gravitational potential.
 
Moreover, multipole acceptance criteria needs be modified for two additional conditions. One is for truncation of long-range interaction and the other is for controlling the accuracy. The count of kernel interactions is determined by the detail of implementation of FMM and MAC. A conventional FMM has a linear stability and \citet{2002JCoPh.179...27D} reports a better stability by using a mass dependent MAC. In this work,  the hybrid method we demonstrate costs more than $\mathcal{O}(N)$ but less than $\mathcal{O}(N\log N)$. The reduction of kernel computation is due to the decoupling the long-range force so that such a hybrid method can robustly work for other kinds of traversal and tree construction as well.

Finally, the current and next generation supercomputers provide a powerful numerical platform to run the massive simulations with unprecedented resolutions and simulation boxes, which usually are composed of tens of thousands of computing nodes and various heterogeneous accelerators and many-core architectures. Besides the pressure of memory, I/O band and storage of snapshot, it requires appropriate algorithms designed for massive concurrency to face the challenge software scalability and computing performance, especially, on the heterogeneous devices. The N-Body applications exchange the enormous number of particles among processes so that the communication strategy becomes an essential issue. The other trend is to employ more efficient method, such as $\mathcal{O}(N)$, to deal with the extreme amount of force computation. The method we proposed provides an option to calculate the force efficiently and decouple the global communication in the meantime. The eventual performance of applications depends on the algorithm and the implementation of programming details. Here, we release a fundamental version of code to verify the  precision and validity of the hybrid algorithm, which is expected to be optimized on high performance computers.

\normalem
\begin{acknowledgements}
We gratefully thank the anonymous referee for the helpful comments. We acknowledge the support from the National Key Program for Science and Technology Research and Development (2017YFB0203300) and the Strategic Priority Research Program of Chinese Academy of Sciences, Grant No. XDC01040100.
\end{acknowledgements}

\bibliographystyle{raa}

\clearpage
\end{document}